\documentclass{elsart3p}

\usepackage{graphicx}
\usepackage{mathrsfs}
\usepackage{amsmath,amssymb}

\begin{document}

\begin{frontmatter}

\title{Quantum reflection of bright matter-wave solitons}
\author{S. L. Cornish\thanksref{label1}\ead{s.l.cornish@durham.ac.uk}}, \author{N. G. Parker\thanksref{label2}}, \author{A. M. Martin\thanksref{label3}}, \author{T. E. Judd\thanksref{label4}}, \author{R. G. Scott\thanksref{label4}},\author{T. M. Fromhold\thanksref{label4}} and \author{C. S. Adams\thanksref{label1}}

\address[label1]{Department of Physics, Durham University, Durham DH1 3LE, UK}
\address[label2]{Department of Physics and Astronomy, McMaster University, Hamilton, Canada}
\address[label3]{School of Physics, University of Melbourne, Parkville,
Victoria 3010, Australia}
\address[label4]{School of Physics and Astronomy, University of Nottingham, University Park, Nottingham NG7 2RD, UK}

\begin{abstract}
We propose the use of bright matter-wave solitons formed from Bose-Einstein condensates with attractive interactions to probe and study quantum reflection from a solid surface at normal incidence. We demonstrate that the presence of attractive interatomic interactions leads to a number of advantages for the study of quantum reflection. The absence of dispersion as the soliton propagates allows precise control of the velocity normal to the surface and for much lower velocities to be achieved. Numerical modelling shows that the robust, self-trapped nature of bright solitons leads to a clean reflection from the surface, limiting the disruption of the density profile and permitting accurate measurements of the reflection probability.
\end{abstract}

\end{frontmatter}

\section{Introduction}
\label{sec:intro}

Solitons are localized self-focusing wavepackets that can propagate over long distances without change in shape, and emerge from collisions
unaltered. Their existence is a common feature of nonlinear wave equations and consequently they are observed in many diverse physical systems including water waves, plasmas, nonlinear optics and particle physics \cite{soliton_book}. Bose--Einstein condensates (BEC) in dilute atomic gases are well-described by a nonlinear Schr\"{o}dinger equation, known as the Gross-Pitaevskii equation (GPE), where the nonlinearity results from the atomic interactions (collisions) in the gas \cite{Dalfovo}. Such systems can support either \emph{dark} solitons \cite{burger,denschlag} or \emph{bright} solitons \cite{Khaykovich,Strecker,Cornish06}, depending on whether the atomic interactions are repulsive or attractive, respectively \cite{gap}. Bright matter-wave solitons manifest themselves as self-trapped condensates where the usual wavepacket dispersion is exactly balanced by the presence of attractive atomic interactions. Their self-trapped nature offers many potential advantages for applications in atom optics and interferometry \cite{interferometry}.

The term \emph{quantum reflection} refers to the process where a particle reflects from a potential without reaching a classical turning point
and is a direct consequence of the wave nature of the particle. Significant reflection occurs when the local wave vector of the particle
$k=\sqrt{(k_\infty^2-2mU(x)/\hbar^2)}$ changes by more than $k$ over a distance of $1/k$, where $k_\infty$ is the wave vector of the particle of
mass $m$ far from the potential $U(x)$. This requires an abrupt variation in the potential $U(x)$, such as can be found in the vicinity of a solid surface. The demonstration of quantum reflection from solid surfaces is typically performed at grazing incidence in order to reduce the wave vector normal to the surface \cite{Anderson,Shimuzu}. The advent of ultracold and quantum degenerate atomic samples with large deBroglie wavelengths opens up new possibilities to study quantum reflection at normal incidence with unprecedented control over the atomic motion. Reflection probabilities as high as $20\%$ have been demonstrated for $^{23}\mbox{Na}$ condensates incident on a solid silicon surface \cite{Pasquini}. More recently, the use of patterned silicon surfaces has resulted in reflection probabilities of $60\%$ \cite{pasquini2full}.

In this paper we propose the use of bright matter-wave solitons to probe and study quantum reflection from a solid surface at normal incidence. In section~\ref{sec:expt} we present details of the proposed experimental scenario and describe how the absence of dispersion enables precise control of the velocity normal to the surface, crucial for the study of quantum reflection. The remainder of the paper is devoted to a numerical study of the quantum reflection of bright matter-wave solitons. In section~\ref{sec:theory} we present the theoretical framework for this investigation. In section~\ref{sec:tanh} we highlight the signatures of quantum reflection by considering the reflection of a soliton from positive and negative step potentials. Finally, in section~\ref{sec:CP} we investigate the quantum reflection from a purely attractive Casimir-Polder atom-surface potential and show that reflection probabilities in excess of $50\%$ are feasible from bulk surfaces.

\section{Experimental Scenario}
\label{sec:expt}

\subsection{Overview}

\begin{figure}[t]
\centering
\includegraphics[width=1.0\columnwidth,clip=true]{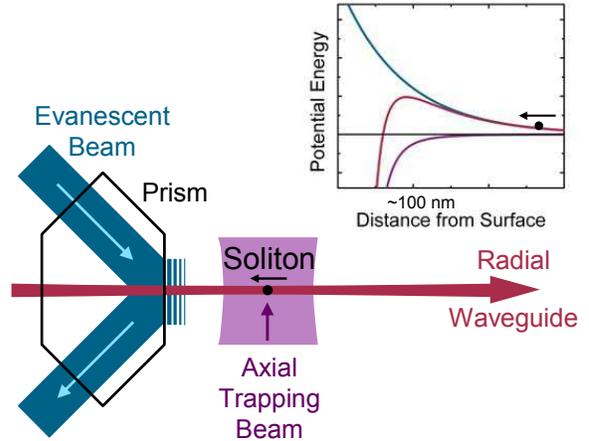}
\caption{Schematic of the proposed experimental configuration for the study of quantum reflection of bright matter-wave solitons. The soliton
propagates towards the surface in an optical waveguide formed by a focussed $1064\,\mbox{nm}$ laser beam. A high power $1030\,\mbox{nm}$ beam intersecting the waveguide at $90^{\circ}$ permits the independent, real time control of the axial trapping. An optional repulsive evanescent field
can be added through the total internal reflection of a $532\,\mbox{nm}$ laser beam in the prism. The inset shows the total potential (red)
experienced by the atoms in the vicinity of the surface as the sum of the Casimir-Polder potential (purple) and the evanescent field (blue).}
\label{fig:surface}
\end{figure}

The proposed experimental scenario for the study of quantum reflection is depicted in Fig.~\ref{fig:surface}. The bright matter-wave soliton will be formed in an optical trap \cite{adams,grimm} from an $^{85}\mbox{Rb}$ condensate \cite{Cornish2000a} with attractive interactions, initially located $\sim 5\,\mbox{mm}$ from the surface of a highly-polished glass prism. Full three dimensional confinement of the initial condensate is provided by two intersecting laser beams \cite{adams}. The first is a focussed $1064\,\mbox{nm}$ beam that forms a radial waveguide with typical harmonic trap frequencies of up to $\sim 100\,\mbox{Hz}$ radially and below $\sim 5\,\mbox{Hz}$ axially. The second beam, from a $50\,\mbox{W}$ $1030\,\mbox{nm}$ laser, is only weakly focussed and intersects the first at $90^{\circ}$. This beam provides the majority of the axial confinement and has no effect on the radial confinement.

This two-beam configuration is chosen so that the harmonic confinement along the waveguide can be altered. In particular, by exploiting the deflection from an acousto-optic device controlling the second beam, the point of intersection of the two beams can be altered leading to real time control of the trap centre along the waveguide. This in combination with the control of the optical power in the second beam allows all parameters of the axial harmonic potential to be manipulated and permits the control of the velocity of the soliton along the waveguide. For example, following the formation of the soliton at the intersection of the two beams, the axial trapping beam can then be shifted in position along the waveguide causing the soliton to be accelerated towards the surface. Removal of the axial trapping potential when the soliton passes through the trap centre leads to a velocity defined by the displacement and strength of the axial trapping potential. The precise control over the parameters of the axial trapping beam afforded by standard optical techniques yields precise control over the velocity of the soliton.

%
%\begin{figure}[t]
%\centering
%\includegraphics[width=1.0\columnwidth,clip=true]{evanescent.eps}
%\caption{The total potential experienced by the atoms in the vicinity of the surface as the sum of the Casimir Polder potential and the
%evanescent field. The height of the resulting repulsive barrier corresponds to the kinetic energy of a soliton moving at $10\,\mbox{mms}^{-1}$
%towards the surface. The inset (not shown yet) shows the ``badlands'' function.} \label{fig:evanescent}
%\end{figure}
%

\subsection{Soliton Formation}

To date, the creation of bright matter-wave solitons \cite{solitary waves} has been demonstrated by three groups at ENS in Paris \cite{Khaykovich}, Rice
University \cite{Strecker} and, more recently, JILA \cite{Cornish06}. All three experiments use a Feshbach resonance \cite{Feshbach} to switch the atomic interactions from repulsive to attractive, triggering the collapse instability of the condensate \cite{Ruprecht,Bradley,Roberts2001a,Donley2001a}, out of which the solitons emerge. The Paris and Rice experiments both used a resonance in the $^{7}\mbox{Li}$ $|\mbox{F}=1,\,\mbox{m}_{\mbox{\scriptsize F}}=1\rangle$ state at $\sim700\,\mbox{G}$, and released the condensates into a quasi-1D optical waveguide in order to demonstrate the solitonic character of the attractive condensates.

Here we consider the use of solitons formed from $^{85}\mbox{Rb}$ condensates in the vicinity of the $\sim 11\,\mbox{G}$ broad Feshbach resonance at $\sim 155\,\mbox{G}$ used in the JILA experiment \cite{Cornish06}. The experiment at JILA demonstrated the controlled creation of both single and multiple solitons and revealed interesting soliton dynamics relevant to the present study. In the case where two solitons were created in the radially symmetric trap, the solitons were observed to oscillate along the weaker axial direction repeatedly colliding in the trap centre. The amplitude of the these oscillations corresponds to a soliton velocity at the trap centre of $\sim 0.5\,\mbox{mms}^{-1}$. This velocity associated with the formation of the solitons for the JILA parameters is of the correct order of magnitude required for the observation of appreciable quantum reflection from a surface \cite{Pasquini}.

\subsection{Atom-Surface Potential and Quantum Reflection}

At distances greater than the atomic scale, the interaction potential of a neutral atom and a solid surface takes the form of an attractive
power-law potential. For distances, $x$, shorter than $\lambda/2\pi$, where $\lambda$ is the wavelength of the dominant atomic transition, the
potential $U(x)=-C_3/x^3$, has the form of a dipole-dipole interaction between the spontaneous atomic dipole and its image
\cite{lenard-jones}. At larger distances the potential becomes $U(x)=-C_4/x^4$ due to retardation of the electrostatic interaction; the
well--known Casimir--Polder result \cite{casimir}.

In the low energy limit $(k\rightarrow0)$, the reflection probability $R$ for a particle incident normally on a surface tends to unity as
$R\simeq1-2\beta_4k$  \cite{Mody,Friedrich}, where $\beta_4$ is the length scale associated with the $C_4$ coefficient,
$C_4=\beta_4^2\hbar^2/2m$. The observation of quantum reflection therefore requires a low incident velocity and a weak attractive force to the
surface. Such conditions were first realized using helium or hydrogen atoms incident on liquid helium surfaces \cite{Nayak,Berkhout} or, for solid surfaces, by working at grazing incidence \cite{Anderson,Shimuzu}. More recently, quantum reflection at normal incidence from a solid surface has been demonstrated using a $^{23}\mbox{Na}$ condensate incident on a silicon surface at very low velocities \cite{Pasquini}. However, in the low velocity limit $(<2\,\mbox{mms}^{-1})$ the large spatial extent of the (repulsive) condensate leads to anomalous reflection, as the front reflects before the tail producing a standing wave that generates
dynamical excitations that disrupt the atom cloud \cite{scottfull,scott-wigner}. Here we demonstrate that the use of well--localized matter--wave solitons will eliminate this problem. This coupled to the precise control of the velocity of the soliton, stands to take the study of quantum reflection to a new level.

\subsection{Evanescent Field}

The experimental configuration also allows for the addition of a repulsive (or attractive) evanescent field in the vicinity of the surface formed by the total internal reflection of a blue (or red) detuned laser field within the glass prism. This produces a potential which decays exponentially with distance from the surface; the decay length being determined by the laser wavelength, the refractive index of the prism and the angle of incidence of the laser beam. When combined with the atom-surface potential, the repulsive evanescent field leads to a repulsive barrier of finite height, in close proximity to the surface (see inset in Fig.~\ref{fig:surface}). Studies of classical reflection from such a barrier can be used to probe the atom-surface potential \cite{Landragin}. Here the low incident velocities of the solitons leads to the peak of the repulsive barrier being located well into the Casimir-Polder regime, unlike in the earlier experiments with atoms dropped from a magneto-optical trap \cite{Landragin}. For example, we estimate that for a soliton incident with a velocity of $10\,\mbox{mms}^{-1}$, setting the barrier height to be equal to the kinetic energy of the soliton leads to the peak being $\sim 170\,\mbox{nm}$ from the surface. Moreover, the addition of both repulsive and attractive evanescent fields can be used to engineer a potential in the vicinity of the surface that significantly enhances the quantum reflection probability \cite{cote}.

\section{Theoretical framework}
\label{sec:theory}

In the limit of ultra-cold temperature the mean-field `wavefunction'
of the BEC $\psi({\bf r},t)$ is well-described by the
Gross-Pitaevskii equation \cite{Dalfovo},
\begin{equation}
i\hbar \frac{\partial \psi}{\partial
t}=\left[-\frac{\hbar^2}{2m}\nabla^2 + V({\bf r})+g|\psi|^2 \right]\psi,
\end{equation}
where $m$ is the atomic mass and the the nonlinear coefficient is
given by $g=4\pi\hbar^2a_{\rm s}/m$ where $a_{\rm s}$ is the {\it
s}-wave scattering length.  The external potential acting on the
system is given by $V({\bf r})$.  This typically includes a trapping
component to confine the atoms $V_T({\bf r})$.  For this we will
assume a cylindrically-symmetric harmonic trap of the form $V_T({\bf
r})=m\omega_r^2 (r^2+\lambda^2 x^2)/2$ where $\omega_r$ is the
radial trap frequency and $\lambda$ characterises the axial trap
strength. Note that the atomic density distribution is related to
the mean-field wavefunction via $n({\bf r},t)=|\psi({\bf r},t)|^2$.

Under very tight radial confinement ($\hbar \omega_r >> \mu$ where $\mu$ is the chemical potential), the wavefunction can be considered ``frozen'' in the radial direction to the non-interacting harmonic oscillator ground state.  By integrating out this dimension, a one-dimensional GPE is obtained in terms of the axial dimension {\it x} with an effective interaction coefficient $g_{\rm 1D}=g/2\pi l_r^2$, where $l_r=\sqrt{\hbar/m\omega_r}$ is the radial harmonic oscillator length.  For $V(x)=0$ the 1D GPE supports exact bright soliton solutions given by,
\begin{equation} \psi(x)=\sqrt{\frac{N}{2\xi}}\,{\rm sech}(x/\xi), \label{eqn:sol} \end{equation}
where $N$ is the number of atoms in the soliton and $\xi=2\hbar/(m|g_{\rm 1D}|N)$ characterises the soliton width.

When the potential is weakly-varying in space, the 1D soliton, to
first order, behaves like a classical particle.  Such classical
behaviour has recently been studied for matter-wave bright solitons
by Martin {\it et al.} \cite{martin}, following analogous particle
approaches in nonlinear optics \cite{aceves}. Quantum reflection,
however, dominates in the opposite regime of large potential
gradients where classical particle analogies are no longer valid. The quantum reflection of matter-wave solitons from a purely attractive potential well has been theoretically studied \cite{leeandbrand}. In that case a sharp switching between the transmission and reflection of the soliton was observed, with bound states of the well playing a key role in the dynamics.

In three-dimensions, the presence of radial confinement
($\omega_r>0$) supports a 3D soliton which is self-trapped in the
axial direction \cite{parker_2007}.  However, in 3D a collapse
instability exists when $N$ exceeds a critical value $N_{\rm c}$ \cite{Ruprecht}.  To avoid collapse effects, we
will restrict our analysis to $N<N_{\rm c}$.  Note that the 3D
soliton may be prone to collapse during its interactions with the
``surface". For example, the interaction of a soliton with an
infinite positive wall is identical to it colliding with another
soliton with a $\pi$-phase difference.  Such collisions have been
shown to exhibit regimes of collapse instabilities
\cite{parker_2008}.

We simulate the interaction of a bright soliton with a surface
potential in both 1D and 3D using the respective GPE.
Time-propagation of the GPE is performed using the Crank-Nicholson
technique \cite{minguzzi}. In 1D the initial soliton solution
corresponds to the analytic soliton solution in
Eq.~(\ref{eqn:sol}), while in 3D it is obtained with a numerical
convergence technique whereby the GPE is run in imaginary time ($t
\rightarrow -it$) from a trial wavefunction \cite{minguzzi}. Note
that when performing 3D simulations we assume cylindrical symmetry
such that the relevant coordinate space is $(x,r)$.  The soliton is
initially positioned a large distance $\Delta x$ to the left of the
surface where the surface potential is negligible. Since we
typically consider an axially-untrapped geometry ($\lambda=0$), a
velocity kick to the soliton is required to induce the dynamics.
This is performed by applying the transformation $\psi({\bf r})
\rightarrow \psi({\bf r})\exp(imvx/\hbar)$, where $v$ is the
required soliton speed. We typically employ velocities in the range
$v=0.1 - 5.0$ mms$^{-1}$. Note that when harmonic trapping is
present in the axial direction, as discussed later, a velocity kick
is not necessary as the soliton freely accelerates towards the
interface.

Following the JILA soliton experiments, we will typically consider a
$^{85}$Rb soliton with radial trap frequency $\omega_r=2\pi
\times 17.5\,\mbox{Hz}$ and axial trapping defined by
$\lambda=0.4$. Furthermore, the scattering length is $a_{\rm
s}=-0.6\,\mbox{nm}$.  For such a system, the critical number is approximately
$N_{\rm c}=2700$ \cite{parker_physica}. We will therefore employ $N
\approx 2000$.

\section{Results 1: Tanh potential}
\label{sec:tanh}

We will first consider a simplified ``surface'' potential to study
the quantum reflection of bright solitons.  We employ a potential
step of the form,
\begin{equation}
V(x)=\frac{V_0}{2}\left[1+{\rm tanh}\frac{x}{\sigma} \right].
\label{eqn:tanh}
\end{equation}
By tuning the height $V_0$ and width $\sigma$ of this smooth
potential, we can control the potential gradient experienced by the
incident soliton. Note that this soliton-step scenario is analogous
to the propagation of optical bright solitons at the interface
between two nonlinear dielectric media of different refractive
indices \cite{aceves}. When the soliton is small compared to the
lengthscale of the step ($\xi << \sigma$), the soliton will behave
similarly to a classical particle. Since we are interested in
studying quantum reflection, we will consider the opposing regime of
sharp tanh potentials where $\sigma \leq \xi$.

It is also interesting to mention analogous studies of dark
matter-wave solitons interacting with step potentials.  These are
localized density dips, supported by repulsive atomic interactions,
and which propagate through an ambient condensate.  This variety of
soliton shows no evidence of quantum reflection when incident on a
negative step potential \cite{parker_2003}.  Here it is likely that
the healing (smoothing) of the ambient condensate across the step
ensures that the dark soliton experiences a smooth effective
potential, thereby preventing quantum reflection.

\subsection{Negative potential $V_0<0$}

\begin{figure}
\centering
\includegraphics[width=1.0\columnwidth,clip=true]{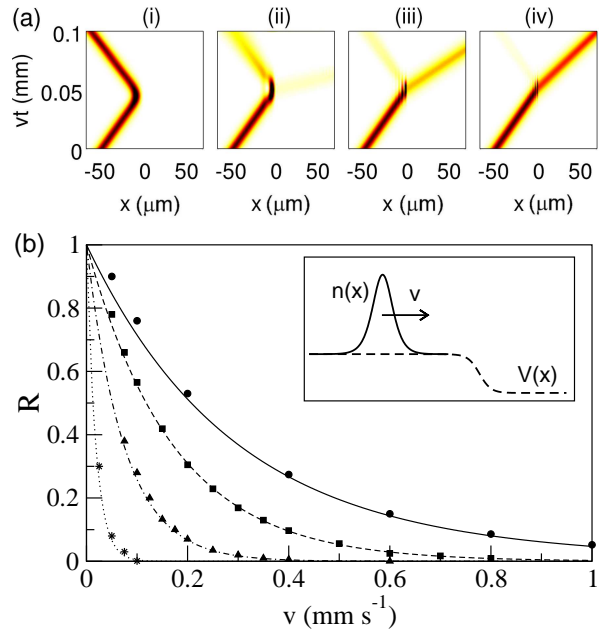}
\includegraphics[width=1.0\columnwidth,clip=true]{negative_tanh2.eps}
\caption{Dynamics of a bright soliton incident on a negative
potential step ($V_0<0$).  (a) Density plots showing a soliton
incident on a hard ($\sigma=0$) negative step for various incident
speeds (i) $v=0.05$, (ii) $v=0.2$, (iii) $v=0.6$ and (iv) $v=1\,\mbox{mms}^{-1}$. (b) Soliton reflection probabilities $R$ as a function of
incident speed $v$ according to the 1D GPE (points) and the
corresponding non-interacting plane wave result (lines). We consider
the negative tanh potential of Eq.~(\ref{eqn:tanh}) with widths of
(from top to bottom) $\sigma/\xi=0$, $0.1$, $0.25$ and $1$. The
inset presents a schematic of the system. The step amplitude is
fixed throughout to be $V_0=-10^{31}$J.} \label{fig:negative}
\end{figure}

We will first consider the regime $V_0<0$, which is illustrated
schematically in Fig.~\ref{fig:negative}(b) (inset).   Classically,
an incident particle would always transmit across such a potential,
gaining speed in the process.

We simulate the impact of the soliton upon the tanh potential using
the 1D GPE.  Assuming the JILA parameters (outlined above) and
$N=1750$, the soliton has a healing length of $\xi=6.4\,\mu$m.  The
radial direction (which is assumed but not explicitly calculated) is
the harmonic oscillator ground state with characteristic width
$l_r\approx 2.6\,\mu$m.  We fix the height of the potential to be
$V_0=-10^{-31}{\rm J} \approx -10\,\hbar \omega_r$ and vary the width
$\sigma$.

Fig.~\ref{fig:negative}(a)
illustrates the dynamics of a soliton incident on a hard negative step potential
($\sigma=0$). At low speed [case(i)], the soliton reflects
elastically from the potential.  Since a classical particle would
transmit across the boundary this is a clear sign of quantum
reflection.  In the opposite regime of high speed [case(iv)], the
soliton transmits over the boundary with negligible perturbation. In
between these limits [cases (ii) and (iii)], we observe a regime where
the soliton splits into a reflected soliton and a transmitted
soliton at the potential. As the speed is increased, we see a
gradual reduction of the reflected soliton and increase of the
transmitted soliton. The reflected soliton typically has the same
speed as the incident soliton, while the transmitted soliton gains
kinetic energy as it passes down the potential and so typically has
larger speed. We also note the appearance of density fringes as the
soliton interacts with the step, with the number of fringes
increasing with the incident speed. Similar fringes appear in the
collisions of two solitons \cite{parker_2008}.

We introduce the reflection probability $R=N_L/N$, where $N_L$ is
the number of atoms to the left of the potential (obtained via
numerical integration) following the interaction.  In
Fig.~\ref{fig:negative} (b) we present the reflection probabilities
$R$ as a function of incident speed $v$ for various widths $\sigma$.
The results of the 1D GPE are shown by points.  Note that the
considerable timescale of simulations at very low speeds limited our
analysis of this regime.

We have also calculated the corresponding reflection probabilities
for a non-interacting plane wave.  This is performed numerically by
dividing the smooth potential into many small intervals, each of
which has constant potential across the interval. The problem is
thereby approximated by a series of sharp finite potential steps,
where the reflection at finite step is given by the standard
analytic solution of the Schr{\"o}dinger equation and the overall
reflection probability is simply the product of these individual
reflection probabilities. These plane wave predictions, shown by
lines in Fig.~\ref{fig:negative} (b), are almost identical to the
GPE predictions. This contrasts strongly with the case of repulsive condensates, where, for $v\lesssim 2\,\mbox{mms}^{-1}$, interatomic interactions cause $R$ to fall significantly below the values predicted for a single atom \cite{Pasquini,scottfull}. Over the range of speeds presented, the kinetic
energy of the soliton is sufficiently large that the interactions
have negligible effect.  We do see small deviations appearing at low
speeds, where the interaction energy of the soliton plays a
significant role.  In this low speed limit, the soliton gives
slightly enhanced quantum reflection over the plane wave solution.

The quantum reflection curves have identical qualitative form,
tending towards unity in the limit of zero speed, and decaying
towards zero as the speed is increased.  Maximum quantum reflection
is seen for a sharp step $\sigma=0$, as expected since the potential
gradient experienced by the soliton is maximal. Accordingly, for
larger widths, the potential gradient is reduced and the reflection
probability decreases.

An important observation is that the reflection probability is very
sensitive to the width of the tanh potential. For example, at speed
$v=0.1\,{\rm mm s}^{-1}$, $R\approx 0.8$ for $\sigma=0$.  For
$\sigma=0.25\xi$ this drops to $R\approx 0.4$ and for $\sigma=\xi$,
the reflection probability is only a few percent.  Such sensitivity
looks promising for the use of matter-wave solitons to probe
atom-surface forces.

\begin{figure}
\centering
\includegraphics[width=1.0\columnwidth,clip=true]{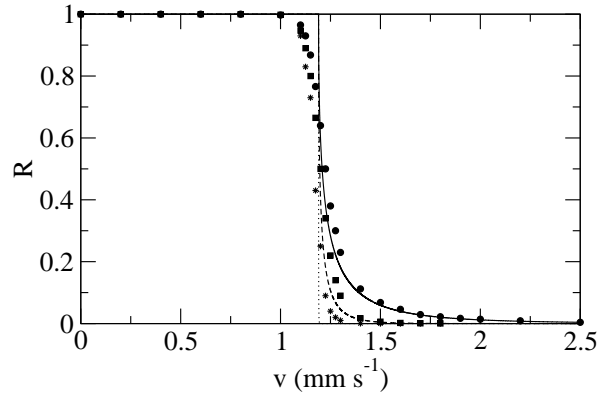}
\caption{Reflection probabilities $R$ for a soliton with speed $v$
incident on a positive tanh step (Eq.~(\ref{eqn:tanh})) according to
the 1D GPE (points) and non-interacting plane-wave approach (lines).
We consider step widths of $\sigma/\xi=0$ (circles/solid line),
$0.1$ (squares/dashed line) and $1$ (stars/dotted line). The step
amplitude is fixed throughout to be $V_0=-10^{31}$J. }
\label{fig:positive}
\end{figure}

\subsection{Positive step $V_0>0$}
We also consider the soliton dynamics when the potential step is positive.  Classically, a particle will transmit over the barrier when $mv^2/2>V_0$, otherwise it reflects.  For $^{85}$Rb atoms and $V_0=10^{-31}$J, the critical speed between reflection/transmission
is $v_{\rm c}\approx 1.17~{\rm mm s}^{-1}$.

Fig.~\ref{fig:positive} presents the reflection probability for a soliton incident on the positive tanh step.  The general features are that the reflection probability is unity at low speed and zero at high speed.  In between there is a narrow region of transition which is centered on the classical critical speed $v_c$. While the GPE results vary smoothly from unity to zero, the plane wave predictions drop suddenly from unity at $v_c$ and decay smoothly for $v>v_c$. This difference is due to the fact that the soliton features a smooth range of velocity (momentum) component, unlike the single-momentum plane wave.  As the step becomes wider, the transition region becomes narrower as the soliton behaves more classically. Recall that for a very wide step $(\sigma>>\xi)$ the soliton would indeed behave like a classical particle and its
reflection curve would be a sharp step function centered at $v=v_c$.

It is important to note that the positive and negative steps give
very different reflection signatures.

\section{Results 2: Casimir-Polder potential}
\label{sec:CP}

We now extend our analysis to a more realistic scenario in which we simulate the quantum reflection of three-dimensional $^{85}$Rb condensates with attractive interactions from a plane silicon surface \cite{silicon}. The condensates approach the surface along the common ($x$) axis of circular symmetry. We use two configurations, a self-trapped condensate with only radial confinement which we call ``soliton $A$'' and an attractive condensate propagating in a full harmonic trap similar to the bright solitary waves observed in the JILA experiment \cite{Cornish06} which we call ``soliton $B$''.

\begin{figure}
\includegraphics[width=1.0\columnwidth]{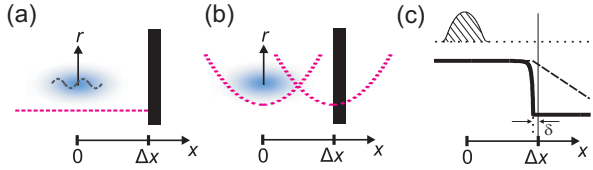}
\caption{\label{fig:scheme}Schematic diagram of the reflection processes and surface potential. (a) and (b) show the initial state for solitons $A$ and $B$ respectively. Blue ovals represent the solitons at their initial position at $x = 0$. Dotted curves represent the longitudinal trapping potentials along the $x$-axis and the black rectangles mark the surface position at $x=\Delta x$. The sinusoidal dashed line in (a) represents the initial phase imprint. The main features of the Casimir-Polder potential are shown in (c). The thick solid line shows the real part of the potential with the dashed line showing the imaginary part. The thin vertical solid line shows the position of the surface at $x = \Delta x$ and the vertical dotted line shows the position of the potential cutoff. The cutoff is a distance $\delta$ from the surface, indicated by arrows. The soliton is represented by the hatched area.}
\end{figure}

Soliton $A$ has 2000 $^{85}$Rb atoms and is confined by radial harmonic trapping only  ($\omega_r=2\pi\times 17.5\,\mbox{Hz}$). Due to the absence of axial trapping, soliton $A$ is only stable due to its attractive inter-atomic interactions which balance out the dispersion force. Soliton $A$ has longitudinal length $\sim 10$ $\mu$m, radial diameter $\sim 5$ $\mu$m and an initial peak density $n_0 \sim 10^{19}$ atoms m$^{-3}$ [Fig. \ref{fig:scheme}(a)]. Soliton $B$ has 1500 $^{85}$Rb atoms in a three-dimensional harmonic trap ($\omega_x=2\pi\times 6.8\,\mbox{Hz}$ and $\lambda=0.4$). Soliton $B$ has longitudinal length $\sim 5$ $\mu$m, radial diameter $\sim 3$ $\mu$m and an initial peak density $n_0 \sim 1.4\times10^{19}$ atoms m$^{-3}$ [Fig. \ref{fig:scheme}(b)].

Initially, the condensate is in its ground state, centred at $x=r=0$ and the silicon surface is positioned at a distance $\Delta x$ away, along the $x$-axis. At time $t=0$ the condensates are suddenly propelled towards the surface. Due to the different axial geometries, we employ different initial conditions depending on whether we are dealing soliton $A$ or $B$, as shown schematically in Fig. \ref{fig:scheme}. For soliton $A$, we numerically apply a velocity kick to the condensate wavefunction at $t=0$ [Fig. \ref{fig:scheme}(a)] as outlined earlier.

For soliton $B$, we instead employ the harmonic trap to set the soliton in motion [Fig. \ref{fig:scheme}(b)], similar to the scheme described in section~\ref{sec:expt}. At $t=0$ we displace the harmonic trap a distance $\Delta x$ along the $x$-axis, so that the centre of the trap coincides with the surface plane. Following the trap displacement each atom in soliton $B$ now has potential energy $V=\frac{1}{2}m\left[\omega_{x}^{2}(x-\Delta x)^2+\omega_{r}^{2}r^{2}\right]$. The displacement causes the soliton to accelerate towards the surface, arriving with a velocity $v \equiv v_x\approx \omega_x\Delta x$. The trap displacement is typically $5 - 30$ $\mu$m, resulting in maximum incident speeds of $0.2 - 1.3$ mm s$^{-1}$. Since the longitudinal radius of soliton $B$ is $\sim 5$ $\mu$m, we cannot perform simulations with $\Delta x < 5$ $\mu$m without introducing additional complications \cite{complications}. We do not have this problem in the case of soliton $A$ \cite{solitonA}.

We now consider the potential due to the Si surface. For a Rb atom at a distance $x' = \Delta x - x$ from a perfectly conducting Si surface, the interaction can be described by the Casimir-Polder potential, $V_{CP}(x')$, which we model with the single-correction function $V_{CP}(x')=-C_4/({x'} ^{3}(x' + 3\lambda_a / 2\pi^2))$ where $C_4=9.1\times10^{-56}$ Jm$^{4}$ and $\lambda_a=780$ nm is the effective atomic transition wavelength for rubidium \cite{Pasquini,casimir,scottfull}. This  potential has no classical turning point and exerts a strong attractive force on the $^{85}$Rb atoms within a distance of $\sim$2 $\mu$m from the surface. Since $V_{CP} \rightarrow -\infty$ as $x\rightarrow \Delta x$ we set $V(x)=V_{CP}(\delta) -  i(x - \Delta x + \delta)V_{im}$ for $x>(\Delta x - \delta)$, where $V_{im}=1.6\times10^{-26}$ Jm$^{-1}$, and $\delta=0.15\mu$m is a small offset from the surface \cite{scottfull}. The imaginary part of the potential models adsorption of Rb atoms by the Si surface. A schematic diagram showing the main features of the model surface potential is shown in
Fig. \ref{fig:scheme}(c).

%Having established the potential landscape for the atoms we
%determine the dynamics of the BECs during the reflection process by
%using the Crank-Nicolson method (cite nr) to solve the
%time-dependent Gross-Pitaevskii equation
%\begin{equation}
%\label{eq:gpe3dcylin}
%-\frac{\hbar^2}{2m}\nabla^2\Psi+V_{T}\Psi+\frac{4\pi\hbar^{2}a}{m}\left|\Psi\right|^2\Psi=i\hbar\frac{\partial\Psi}{\partial t}
%\end{equation}
%where $a=-0.6$ nm is the $s$-wave scattering length, $\nabla^2$ is
%the Laplacian in cylindrical coordinates, $\Psi(x,r,t)$ is the wave
%function at time $t$, normalized such that $\left|\Psi\right|^2$ is
%the number of atoms per unit volume, and

\begin{figure}
\includegraphics[width=1.0\columnwidth]{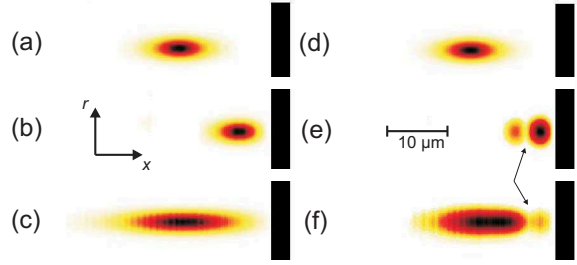}
\caption{\label{fig:snaps}Density slices (dark shading indicates high density) in $x$-$r$ plane (axes inset) for 3D $^{85}$Rb condensates with only radial confinement reflecting from a planar Si surface at incident velocities $v = 0.1$ mm s$^{-1}$ (left, panels (a),(b) and (c)) and $v = 0.4$ mm s$^{-1}$ (right, panels (d),(e) and (f)). The images for the lower velocity case are taken at time $t=0$ ms (a), $152$ ms (b) and $300$ ms (c). The images for the higher velocity case are taken at time $t=0$ ms (d), $38$ ms (e) and $75$ ms (f). Small arrows indicate the position of density nodes - the colour scale in (e) and (f) has been adjusted to highlight these. Black rectangles show the position of the surface.}
\end{figure}

We now consider the dynamics of the reflection process under the combined potential $V = V_{CP} + V_T$. Fig.~\ref{fig:snaps} shows density slices of soliton $A$ before, during and after quantum reflection from the Si surface at two different velocities. The left panels ((a), (b) and (c)) show the reflection process for an incident velocity $v = 0.1$ mm s$^{-1}$ and the right panels ((d), (e) and (f)) show the reflection process for an incident velocity $v = 0.4$ mm s$^{-1}$.

First we analyse the lower velocity case. Fig.~\ref{fig:snaps}(a) shows the equilibrium density profile of soliton $A$ at the start of the simulation and its position relative to the surface. We see that the condensate has a ``cigar'' shaped profile. Fig.~\ref{fig:snaps}(b) shows soliton $A$ 152 ms into the simulation. The soliton is just starting to move away from the surface having reflected. The soliton has been slightly compressed but we see no major density modulations because the de Broglie wavelength is longer than the soliton's longitudinal diameter. Fig.~\ref{fig:snaps}(c) shows soliton $A$ after 300 ms. The condensate is now considerably more elongated. This is because the soliton has lost many atoms ($\sim 50\%$) and hence its axial size, which
scales like $1/N$, grows considerably.  The reflection process may also excite shape excitations in the outgoing soliton.  If such excitations
are sufficiently large, kinetic effects may overcome the attractive interactions and lead to the formation of a dispersive wavepacket, as
predicted by a variational approach \cite{parker_2007}.  Note that a change in the scattering length following the reflection could be used to preserve the
solitonic nature of the condensate. The small modulations observed in the density are propagating sound waves. Note that the condensate's density profile has no major nodes or topological excitations following reflection. This is in stark contrast with the low velocity quantum reflection of condensates with repulsive interactions \cite{scottfull} where lobes and vortices were seen to form. For the condensates considered here, the attractive interactions resist the formation of such features because low density regions have higher mean-field energies. In this case, the condensate has a low kinetic energy ($\sim 7\times10^{-34}$J) compared with the interaction energy ($\sim 6\times10^{-33}$J) so interaction effects dominate the condensate's behaviour.

We now analyse the higher velocity case for soliton $A$. The soliton starts as before [Fig. \ref{fig:snaps}(d)]. At 38 ms the soliton is just starting to move away from the surface [Fig. \ref{fig:snaps}(e)]. A density node has formed in the cloud (indicated by small arrows) because the de Broglie wavelength, $\sim 10$ $\mu$m, is similar to the longitudinal diameter of the condensate and the condensate's kinetic energy is sufficient to overcome the interactions' resistance to sharp features. Interestingly, the node is very well defined when compared with the non-interacting and repulsive interaction cases \cite{scottfull}. It extends all the way to zero  density and leaves a comparatively large volume where we do not expect to find many atoms. This is due to the mean field attractive interactions which naturally pull atoms away from energetically unfavourable lower density areas. Fig.~\ref{fig:snaps}(f) shows the condensate's density after reflection at $t=75$ ms. Again we observe an elongation of the cloud because the condensate has lost atoms, and hence its ability to self-trap. As in the low velocity case, the calculations reveal no vortices or topological excitations but do show modulations due to propagating sound waves. The density node, formed during reflection, is still present in the density profile (indicated by small arrows). Although the interactions initially resist such features, once there is enough kinetic energy to form them, this configuration is stable.

There is a possibility that reflection processes can trigger a collapse in attractive condensates if the spatial compression at the surface causes the peak density to rise above a certain value \cite{parker_2007}. Although the peak density in these calculations does increase by a factor of up to $1.5$, we see that this never triggers collapse for the parameters considered here. By analogy to collisions of solitons with $\pi$ phase differences we would expect collapse instabilities to occur for intermediate speeds and for solitons that are themselves closer to collapse \cite{parker_2008}.

\begin{figure}
\includegraphics[width=1.0\columnwidth]{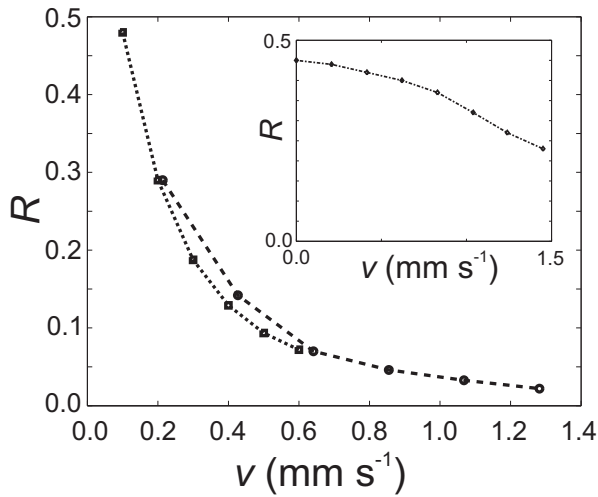}
\caption{\label{fig:refprobs}Reflection probabilities, $R$, plotted
against incident velocity, $v$, for 3D $^{85}$Rb condensates
reflecting from a planar Si surface. The dotted curve with square
points is for $^{85}$Rb BEC $A$ which only has radial confinement
and the dashed curve with circular points is for the $^{85}$Rb BEC
$B$ with in a 3D harmonic trap. The inset shows $R(v)$ for a $^{23}$Na
BEC with repulsive interactions (see text for details).}
\end{figure}

Fig.~\ref{fig:refprobs} shows how the reflection probabilities, $R$, of soliton $A$ (dotted line, square points) and soliton $B$ (dashed line, circular points) vary with incident velocity $v$. Both curves show the qualitative trend expected for a non-interacting wave packet i.e. $R$ decreases with increasing $v$. The different propagation methods have little influence on the reflection probability and the two curves approximately overlap. As noted before, the phase imprint method makes it straightforward to use lower velocities, hence we can reach $0.1$ mm s$^{-1}$ without introducing any complications associated with the initial state touching the surface. For this low velocity we predict $R = 0.48$. This is comparable with the highest reflection probabilities observed in experiments and simulations with sodium atoms in spite of the fact that the rubidium atom is almost four times heavier and no nanofabrication technique has been used to enhance the reflectivity of the surface \cite{scottfull,pasquini2full}. This is primarily because lower velocities can be readily attained using attractive BECs due to their small spatial extent and ability to self-trap.

We now make a simple comparison between quantum reflection of condensates with  attractive interactions and condensates with repulsive interactions. The inset of Fig. \ref{fig:refprobs} shows $R$ plotted against $v$ for simulations of a $^{23}$Na BEC, which is taken to be one-dimensional in the $x$-direction. The trap frequency is $\omega_x=2\pi\times 3.3\,\mbox{Hz}$ and the wavefunction is normalised such that $n_0 = 2.2\times10^{18}$ atoms m$^{-3}$ as in recent experiments \cite{Pasquini}. Repulsive condensates do not self-trap, hence we use the trap displacement method to propel the BEC along the $x$-axis towards the surface. We model the reflection process using the 1D Gross-Pitaevskii equation with the Casimir-Polder potential assuming $a_{\rm s}=+2.9\,\mbox{nm}$ and $\lambda_a = 590$ nm.

We immediately see that unlike the attractive case, reflection probabilities do not continue to rise as $v \rightarrow 0$, but rather saturate at a maximum value, $\sim 0.45$. This occurs because, at low $v$, the condensate spends a long time in contact with the surface where the curvature of the wavefunction is high. Combined with the repulsive nature of the interactions, this creates a quantum pressure which drives atoms onto the surface. This problem is particularly acute for the repulsive condensates we consider here because the trap displacements required to attain low $v$ are lower than the longitudinal diameter of the condensate which therefore starts in contact with the surface and remains in contact throughout the simulation. An attractive condensate in the form of a soliton, on the other hand, prefers to avoid contact with the surface. It is, in fact, impossible to create an attractive condensate ground state with significant initial surface contact because the wavefunction finds sharp features energetically unfavourable and subsequently moves away from the surface to avoid creating high levels of quantum pressure. Attractive condensates have a further advantage in that they also move away from absorbing potentials - it is possible to reflect an attractive condensate from a purely absorbing imaginary potential which has no variation of its real part. This is because the absorbing region of the potential draws atoms out of the condensate, thereby raising its energy and causing the condensate to move away from this region.

%We conclude that BECs with attractive interactions have a number of advantages in terms of quantum reflection. Self-trapped BECs can be propagated towards surfaces at much lower velocities and they shy away from both real and absorbing potentials. In addition they tend to retain their form during the reflection process which limits disruption although this may restrict their usefulness if one wishes to manipulate the phase and density of the cloud.

\section{Conclusions}
\label{sec:conclusion}

In conclusion we have presented details of a proposal to use bright matter-wave solitons to probe and study quantum reflection from a solid surface at normal incidence. We have demonstrated that condensates with attractive interactions have a number of advantages in terms of quantum reflection. The absence of dispersion as the soliton propagates allows for precise control of the velocity normal to the surface and for much lower velocities to be achieved. Our numerical results show that the robust, self-trapped nature of bright solitons leads to a clean reflection from the surface, limiting the disruption of the density profile. This permits accurate measurements of the reflection probability which can easily exceed $50\%$ and offers a new method to probe the atom-surface potential. Such sensitive measurements in close proximity to a massive object may, in the future, present a new method to test the short range corrections to gravity due to exotic forces beyond the Standard model \cite{dimopoulos}.

\ack We thank J. Brand for many stimulating discussions. We acknowledge support from the UK EPSRC, the Royal
Society (SLC), the Canadian Commonwealth Scholarship Program (NGP) and the Australian Research Council (AMM).

%%%%%%%%%%%%%%%%%%%%%%%%%%%%%%%%%%%%%%%%%%%% BIBLIOGGRAPHY %%%%%%%%%%%%%%%%%%%%%%%%%%%%%%%%%%%%%%%%%%%%%%%%%%%%%%%%%%%%%%%%%%%%%%%%%%%%%%%%%%%%%%

\end{document}